\documentclass[10pt,twoside,twocolumn,a4paper]{article}

\usepackage[accepted]{bpasts}

\usepackage{t1enc}
\usepackage[utf8]{inputenc}
\usepackage{amsmath,amsfonts}

\usepackage{graphicx}
\usepackage{flushend}

\usepackage{amsmath}
\usepackage{amsfonts}
\usepackage{dsfont}
\usepackage{hyperref}

\newtheorem{prop}{Proposition}

\newcommand{\Expectation}{\ensuremath{\mathds{E}}}
\newcommand{\Wg}{W\hspace{-0.7mm}g}
\newcommand{\eg}{\emph{e.g.}}
\newcommand{\ie}{\emph{i.e.}}
\newcommand{\kk}{\mathbf{k}}
\newcommand{\U}{\mathrm{U}}
\newcommand{\tr}{\mathrm{tr}}
\newcommand{\1}{{\rm 1\hspace{-0.9mm}l}}
\newcommand{\halmos}{\hfill$\square$}
\newcommand{\IntU}{\texttt{IntU}}
\newcommand{\Mathematica}{\emph{Mathematica}}

\usepackage{setspace}

\newcommand{\IncludePdfExample}[1]{\\[0.15cm] \includegraphics[scale=1]{#1} \newline}

\abtitle{Symbolic integration with respect to the Haar measure on the unitary group}
\title{Symbolic integration with respect to the Haar measure on the unitary groups}

\abauthor{Z. Pucha{\l}a, J. A. Miszczak}
\author{Z. PUCHA{\L}A$^{1} $\email{z.puchala@iitis.pl} , J. A. MISZCZAK$^1$}

\Address{$^1$ Institute of Theoretical and Applied Informatics, Polish Academy
of Sciences, Ba{\l}tycka~5, 44-100~Gliwice, Poland}

\Abstract{We present \IntU{} package for \Mathematica{} computer algebra system. The presented package  performs a symbolic integration of polynomial functions over the unitary group with respect to unique normalized Haar measure.  We describe a number of special cases which can be used to optimize the calculation speed for some classes of integrals. We also provide some examples of usage of the presented package.}

\vol{61} \no{1} \year{2017}
\doi{10.1515/bpasts-2017-0003}

\begin{document}
\maketitle

\section{Introduction}
The integration over unitary group is an important subject of studies in many
areas of science, including mathematical physics, random matrix theory and
algebraic combinatorics. In 2006 Collins and \'Sniady
\cite{collins06integration} proved a formula for calculating  monomial integrals
with respect to the Haar measure on the Unitary group
\begin{equation} \label{eqn:the-interal}
\int_{\U(d)} U_{IJ} \overline{U}_{I'J'} dU = \int_{\U(d)}  U_{i_1 j_1} \dots U_{i_n j_n} 
                 \overline{U}_{i'_1 j'_1} \dots \overline{U}_{i'_n j'_n} dU. 
\end{equation}
Integrals of the above type are known as \emph{moments of the $\U(d)$} and are
well-known in mathematical physics literature for a long time.
The problem of the integration of elements of unitary matrices was for the first
time considered in the context of nuclear physics in  \cite{ullah63expectation}.
The asymptotic behaviour of the integrals of the type (\ref{eqn:the-interal}) was 
considered by Weingarten in~\cite{weingarten78asymptotic}.

In this paper we describe a \Mathematica{} package \IntU{}~\cite{intu} for calculating
polynomial integrals over  $\U(d)$ with respect to the Haar measure. We describe a
number of special cases which can be used to optimize the calculation  speed for
some classes of integrals. We also provide some examples of usage of the presented
package including the applications in the study of the geometry of the quantum
states.

%

This paper is organised as follows.
In Section~\ref{sec:math-background} we introduce notation present mathematical
background concerning polynomial integrals over unitary group.
In Section~\ref{sec:special-cases} we describe some special cases, in which the 
integration can be calculated more efficiently. 
In Section~\ref{sec:package-desc} we provide the description of the \IntU{} package
with the list of main functions.
In Section~\ref{sec:examples} we show some examples of the usage.
In Section~\ref{sec:summary} we provide a summary of the presented results
and give conclusions.

\section{Mathematical background} \label{sec:math-background}

\subsection{Basic concepts} \label{sec:basic}

We denote by ${M}_{n}$ square matrices of size $n$.
The compact group of $d \times d$ unitary matrices we denote as $\U(d)$. We
equip the above group with unique normalized Haar measure denoted by $dU$.
Random elements distributed with measure $dU$ form so called \emph{Circular
Unitary Ensemble}. 

Integer partition $\lambda$ of a positive integer $n$ is a weakly decreasing
sequence $\lambda = (\lambda_1, \lambda_2, \dots , \lambda_l)$ of positive
integers, such that $\sum_{i=1}^{l} \lambda_i =|\lambda|= n$. To denote that
$\lambda$ is a partition of $n$ we write $\lambda \vdash n$. The length of
a partition is denoted by $l(\lambda)$.
By $\lambda \sqcup  \mu$ we denote a partition of $n_1+n_2$ obtained by joining 
partitions $\lambda \vdash n_1$ and $\mu \vdash n_2$.

Each permutation $\sigma \in S_n$ can be uniquely decomposed into a sum of
disjoint cycles where the lengths of the cycles sum up to $n$.  Thus the vector of the
lengths of the cycles, after reordering, forms a partition $\lambda \vdash n$.
The partition $\lambda$ is called the cycle type of permutation $\sigma$.

\subsection{Moments of the $\U(d)$}
Let us consider a polynomial $p$. From the linearity of an integral we have 
\begin{equation}
 \int_{\U(d)} p(U) dU = \sum_{I,J,I',J'} c(I,J,I',J') \int_{\U(d)} U_{IJ}
 \overline{U}_{I'J'} dU,
\end{equation}
where $I,J,I',J'$ are multi-indices and $c$ are the coefficients of $p$. 
The value of such monomial integrals is given as \cite{collins06integration}
\begin{equation}
\begin{split} \label{intFormula}
\int_{\U(d)} U_{IJ} \overline{U}_{I'J'} dU 
= \int_{\U(d)}  U_{i_1 j_1} \dots U_{i_n j_n} 
                 \overline{U}_{i'_1 j'_1} \dots \overline{U}_{i'_n j'_n} dU = 
 \\ 
 =
 \sum_{\sigma, \tau \in S_n} \delta_{i_1,i'_{\sigma(1)}} \dots 
 \delta_{i_n,i'_{\sigma(n)}} \delta_{j_1,j'_{\tau(1)}}  \dots 
 \delta_{j_n,j'_{\tau(n)}} 
 \Wg(\tau \sigma^{-1}, d),
\end{split}
\end{equation}
where $\Wg $  is the Weingarten function discussed below. 
In the case where the multi-indices differ in length, \ie{} $n \neq n'$, we have
\begin{equation}
 \int_{\U(d)} U_{IJ} \overline{U}_{I'J'} dU  = \int_{\U(d)}  U_{i_1 j_1} \dots U_{i_n j_n} \overline{U}_{i'_1 j'_1} \dots
\overline{U}_{i'_{n'} j'_{n'}} dU = 0.
\end{equation}
The integrals of the above type are known as \emph{moments of the $\U(d)$}.

\subsection{Weingarten function} \label{sec:weingarten}
The \emph{Weingarten function} \cite{collins06integration}
is defined for $\sigma\in S_n$ and positive integer $d$, as
\begin{equation} \label{eqn:weingarten-perm-definition}
 \Wg(\sigma, d) = \frac{1}{(n!)^2}
 \sum_{\genfrac{}{}{0pt}{}{\lambda \vdash n}{l(\lambda) \leq d}} 
 \frac{\chi^{\lambda}(e)^2}{s_{\lambda,d}(1)} \chi^{\lambda}(\sigma),
\end{equation}
where the sum is taken over all integer partitions of $n$ with length
$l(\lambda) \leq d$, $s_{\lambda,d}(1)$ is the Schur polynomial $s_{\lambda}$
evaluated at $(\underbrace{1,1,\dots,1}_{d})$ and $\chi^{\lambda}$ is an
irreducible character of the symmetric group $S_n$ indexed by partition
$\lambda$.

\subsubsection{The dimension of irreducible representation of $\U(d)$}

The value of the Schur polynomial at the point $(\underbrace{1,1, \dots
,1}_{d})$, \ie{} the dimension of irreducible representation of $\U(d)$
corresponding to partition $\lambda$, is equal to (see \eg{} \cite[Theorem
6.3]{fultonharris})
\begin{equation} \label{eqn:schurAt1}
s_{\lambda,d}(1) =  s_{\lambda}(\underbrace{1,1, \dots ,1}_{d}) = 
 \prod_{1 \leq i < j \leq d} \frac{\lambda_i - \lambda_j + j - i }{j - i}.
\end{equation}
%

\subsubsection{Irreducible character of $S_n$} \label{sec:characterSn}

The irreducible character of $S_n$ indexed by partition $\lambda$, 
$\chi^{\lambda}(\sigma)$ depends on a conjugacy class of permutation $\sigma$.
Two permutations are in the same conjugacy class if and only if they have the
same cycle type, thus it is common to write  $\chi^{\lambda}(\sigma) =
\chi^{\lambda}(\mu)$, where $\mu$ is an integer  partition corresponding to the
cycle type of $\sigma$.

In the case of identity permutation the cycle type is given by a trivial
partition, $e = \{\underbrace{1,1,\dots,1}_{n} \}$, and the character value is
equal to the dimension of the irreducible representation of $S_n$ indexed by
$\lambda$. In this case it is given by the celebrated \emph{hook length} formula
\cite[Eq.~4.12]{fultonharris}
\begin{equation} \label{eqn:hook}
 \chi^{\lambda}(e) = \frac{|\lambda|! }{ \prod_{i,j} h_{i,j}^{\lambda}},
\end{equation}
%
%
where $|\lambda| = \lambda_1 + \lambda_2 + \dots + \lambda_{l(\lambda)}$,  and
$h_{i,j}^{\lambda}$ is the hook length of the cell $(i,j)$ in a Ferrers diagram
corresponding to partition $\lambda$, see \eg{}
\cite[p.~57]{james1981representation}.

In the case of a non-trivial partition the character of symmetric group 
$\chi^{\lambda}(\sigma) = \chi^{\lambda}(\mu)$ can be evaluated with the use of
Murnaghan-Nakayama rule (see \eg{} \cite[Th. 4.10.2]{sagan2001symmetric}), which
describes a combinatorial way of calculating the character. 
\cite{bernstein2004computational} is used. 

From the above considerations one can notice that the Weingarten function
depends only on a cycle type of a permutation $\sigma$ and thus it is constant
on a conjugacy class represented by $\sigma$. Thus we may define the Weingarten
function as 
\begin{equation}\label{eqn:weingarten-type-definition}
 \Wg(\mu, d) = \frac{1}{(|\mu|!)^2} \sum_{\genfrac{}{}{0pt}{}{\lambda \vdash
|\mu|}{l(\lambda) \leq d} }
 \frac{\chi^{\lambda}(e)^2}{s_{\lambda,d}(1)} \chi^{\lambda}(\mu),
\end{equation}
where $\mu$ is an integer partition, which is a cycle type of $\sigma$.

\section{Special cases}\label{sec:special-cases}

In this section we present some special cases of integrals, with respect to  the
Haar measure on the unitary group. In these cases the value of the integral can
be calculated without the direct usage of formula~(\ref{intFormula}), which requires
processing of $ \prod_i^d k_i !  \prod_j^d l_j !$ permutations, where $k_i \
(l_j)$ denotes the number of $i  \ (j)$ in multi-indices $I \ (J)$ respectively.

The presented special cases have been implemented in the package, to increase
its efficiency. This goal has been achieved by minimizing the number of calculations
of the Weingarten function.

\subsection{First two moments of $\U(d)$}
In the case of monomials of rank equal to $2$ and $4$, we have 
\cite[Prop. 4.2.3]{hiai2006semicircle} formulas
\begin{equation}
\label{eqn:int-formula-1-idx} 
\int_{\U\left(  d\right)} u_{i j} \overline{u}_{i^{\prime} j^{\prime}} dU  
=\frac{1}{d} \delta_{i i^{\prime}}  \delta_{j j^{\prime}},
\end{equation}
and
\begin{multline}
 \label{eqn:int-formula-2-idx}
\int_{\U \left(  d\right)} u_{i_{1} j_{1}}u_{i_{2} j_{2}}\overline
{u}_{i_{1}^{\prime} j_{1}^{\prime}}\overline{u}_{i_{2}^{\prime} j_{2}^{\prime}}dU
= \\
=\frac{
	\delta_{i_{1} i_{1}^{\prime}}
	\delta_{i_{2} i_{2}^{\prime}}
	\delta_{j_{1} j_{1}^{\prime}}
	\delta_{j_{2} j_{2}^{\prime}}
+   \delta_{i_{1} i_{2}^{\prime}}
	\delta_{i_{2} i_{1}^{\prime}}
    \delta_{j_{1} j_{2}^{\prime}}
    \delta_{j_{2} j_{1}^{\prime}}
 }{d^{2}-1} +
\\ 
- 
\frac{  
	\delta_{i_{1} i_{1}^{\prime}}
	\delta_{i_{2} i_{2}^{\prime}}  
	\delta_{j_{1} j_{2}^{\prime}}
	\delta_{j_{2} j_{1}^{\prime}}
+   \delta_{i_{1} i_{2}^{\prime}}
	\delta_{i_{2} i_{1}^{\prime}}
    \delta_{j_{1} j_{1}^{\prime}}
    \delta_{j_{2} j_{2}^{\prime}}
}
{d\left( d^{2}-1\right)}
,
\end{multline}
allowing calculation of the values for polynomial integrals of degree less
than~$5$ without the direct usage of the formula~(\ref{intFormula}). 

This optimization gives only a minor improvement of efficiency, as in these
cases the direct calculation of Weingarten function is very fast. 

\subsection{Elements from one row (column)} \label{sec:special-1-row}
The next optimization is based on the fact that the distribution of random
vector consisting of squares of absolute values of a row (or a column) of
unitary matrix  distributed with the Haar measure, is uniform on a standard
$d$-simplex $\Delta^d$~\cite[Ch.~4]{hiai2006semicircle}
\begin{equation}
\{|u_{i,1}|^2, |u_{i,2}|^2, \dots, |u_{i,d}|^2 \} \sim \mathbb{U}(\Delta^d),
\end{equation}
 where $\mathbb{U}(A)$ denotes the normalized uniform measure (proportional to
Lebesgue measure) on a set $A\subset \mathbb{R}^d$, and standard $d$-simplex
$\Delta^d$ is defined as, $ \Delta^d = \{\lambda \in \mathbb{R}^d:\lambda_i \geq
0 , \sum_{i=1}^d \lambda_i = 1 \}.$ 

Using Beta integral, one obtains that for a fixed row $i_0$ and a vector $p$
with non-negative entries $p_j$, we have the following 
\begin{equation}
 \int_{\U(d)} \prod_{j=1}^d |u_{i_0,j}|^{2 p_j} dU =  
  \Gamma(d) \frac{\Gamma(p_1+1) \times \dots \times \Gamma(p_d+1) }
       {\Gamma(p_1 +  \dots + p_d + d)}.
\end{equation}
The above, as a special case, gives us:
\begin{eqnarray} 
 \int_{\U(d)} |u_{i_0,j}|^{2 k} dU &=& \frac{(d-1)! k!}{(d-1+k)!}, \\
 \int_{\U(d)} |u_{i_0,j}|^{2} |u_{i_0,k}|^{2} dU &=& \frac{1}{d (d+1)},
\end{eqnarray}
which can be found in literature \cite{hiai2006semicircle,donati10truncations}.

This optimization allows for an enormous improvement in efficiency thanks to
avoiding  $(\sum p_i) ! \prod p_i !$ executions of Weingarten function needed in
the case  of the direct usage of formula (\ref{intFormula}).

\subsection{Even powers of diagonal element absolute values}
\label{sec:even-powers}

In this subsection we consider the integrals of the type 
\begin{equation}
 \int_{\U(d)} |u_{i,j}|^{2 p} |u_{k,l}|^{2 q} dU,
\end{equation}
where $p,q$ are non-negative integers and $i \neq k$, $j \neq l$.  In the case
of  $p=q=1$ this integral is known \cite[Prop. 4.2.3]{hiai2006semicircle}
\begin{equation}
 \int_{\U(d)} |u_{i,j}|^{2} |u_{k,l}|^{2} dU = \frac{1}{d^2 - 1}.
\end{equation}
For general non-negative integers $p,q$ the following proposition is true.
\begin{prop}
Let $p,q$ be non-negative integers, for  $i \neq k$ and  $j \neq l$, we have
\begin{equation} \label{newIntegral}
 \int_{\U(d)} |u_{i,j}|^{2 p} |u_{k,l}|^{2 q} dU  =  
   p! q! \sum_{\genfrac{}{}{0pt}{}{\lambda \vdash p}{\mu \vdash q}} 
          \kk_{\lambda}  \kk_{\mu}\Wg( \lambda \sqcup \mu, d),
\end{equation}
where the above sum is taken over all integer partitions of $p$ and $q$. Symbol 
$\kk_{\nu}$ denotes a cardinality of conjugacy class for a permutation with 
cycle type given by partition $\nu \vdash r$ \cite[Eq. 1.2]{sagan2001symmetric}
\begin{equation} \label{eqn:k-lambda}
\kk_{\nu}  = \frac{r!}{1^{m_1} m_1 ! 2^{m_2} m_2 ! \cdots r^{m_r} m_{r} !} ,
\end{equation}
where $m_i $ denotes the number of $i$ in partition $\nu$.
\end{prop}
The above is a special case of more general fact.
\begin{prop} \label{prop:special-even}
For any permutation $\pi \in S_d$ and any non-negative integers 
$p_1, p_2, \dots p_d$, we have
\begin{equation} \label{new-int}
\begin{split}
 \int_{\U(d)} \prod_{j=1}^d |u_{j,\pi(j)}|^{2 p_j}dU &= 
 \int_{\U(d)} \prod_{j=1}^d |u_{j,j}|^{2 p_j}dU =
 \\ 
   =
 \left(\prod_{j=1}^d p_j ! \right)&
   \sum_{\lambda_1 \vdash p_1} \sum_{\lambda_2 \vdash p_2} \dots 
   \sum_{\lambda_d \vdash p_d}
   \left(\prod_{j=1}^d \kk_{\lambda_j} \right)\times \\\times
   &\Wg ( \lambda_1 \sqcup \lambda_2 \sqcup \dots \sqcup \lambda_d , d).
   \end{split}
\end{equation}
\end{prop}
\proof  If we apply the formula from Eq.~(\ref{intFormula}) to integral
(\ref{new-int}), then the non-vanishing permutations of indices are these
which permute within the blocks of sizes $\{p_1,p_2,\dots, p_d\}$, \ie{}
$$
\sigma =  \sigma_1 \oplus \sigma_2 \oplus \dots \oplus \sigma_d,
$$
where $\sigma_j \in S_{p_j}$ and it permutes indices in $j^{\text{th}}$ block of
size $p_j$. The same situation holds in the case of the second indices. Thus the
permutation $\tau\sigma^{-1}$ is also in this form. Each permutation of the
above type will be present in the  sum $\prod_{j=1}^d p_j !$ times. The cycle
type of such permutations is given by a partition which  is obtained by joining
cycle types of small permutations. Since Weingarten function depends only on a
cycle type of permutation, the size of conjugacy class is calculated for each
partition, instead of evaluating multiple times Weingarten function. \halmos

The formula (\ref{new-int}) is far less computationally demanding than the
direct usage of (\ref{intFormula}). One can notice that
Proposition~\ref{prop:special-even} allows us to avoid at least  $\prod_{j=1}^d
p_j!$ executions of Weingarten function comparing to the direct usage of
(\ref{intFormula}). However, the exact time-efficiency depends also on the
cardinality of conjugacy classes for partitions of $p_1, \dots , p_d$.

\subsection{Cycle permutations}

In this section we consider another special type of integral for positive
integer $k$ and a permutation $\sigma$ of $\{1,2,\dots,m\}$ being a cycle
\begin{equation}
 \int_{\U(d)} \left( \prod_{i=1}^m u_{i,i} \overline{u}_{i,\sigma(i)} \right)^k d U.
\end{equation}
We have the following proposition.
\begin{prop}
Let $m \leq d$, $k$ is a positive integer and $\sigma \in S_m$ be a cycle, \ie{}
the cycle type is given by partition $\{m\}$. Then

\begin{equation} \label{eqn:cycle-opt}
\int_{\U(d)} \left( \prod_{i=1}^m u_{i,i} \overline{u}_{i,\sigma(i)} \right)^k d U
= (k!)^{2m - 1} \sum_{\lambda \vdash k } \kk_{\lambda} \Wg( m \lambda , d).
\end{equation}
\end{prop}
\proof  One can notice that cycle lengths of permutations in this setting  must
be divisible by $m$. The number of permutations with cycle type given by a
particular partition can be easily obtained by the usual counting argument.
\halmos

Using the above formula (\ref{eqn:cycle-opt}) one obtains an enormous
improvement  in efficiency by avoiding more than $(k!)^{2m - 1}$ executions of
Weingarten function, comparing to the case of the direct usage of
(\ref{intFormula}).
%

\section{Package description}\label{sec:package-desc}
Below we describe the functions implemented in \IntU{} package. The functions
are grouped in three categories: functions implementing the main functionality,
functions related to the calculation of Weingarten function and helper
functions.

\subsection{Main functionality}

The main functionalty of \IntU{} package is provided be the
\texttt{IntegrateUnitaryHaar} and \texttt{IntegrateUnitaryHaarIndices}
functions. The first one operates directly on polynomial expressions, while the
second one accepts four-tuple of indices. The examples of the usage are given in
Section~\ref{sec:examples}.

\begin{itemize}
 \item \texttt{IntegrateUnitaryHaar[integrand,\{var,dim\}]} --
 gives the definite integral on unitary group with respect to the Haar measure, 
 accepting the following arguments
 \begin{itemize}
 	\item \texttt{integrand} -- the polynomial type expression of variable
	\texttt{var} with indices placed as subscripts, can contain any other
	   symbolic expression of other variables,   
	\item \texttt{var} -- the symbol of variable for integration,   
	\item \texttt{dim} -- the dimension of a unitary group, must be a positive integer.
 \end{itemize}
This function is presented in the examples described in Sections
\ref{ex:elementary}, \ref{ex:matrix} and \ref{ex:moments}.  

\item \texttt{IntegrateUnitaryHaar[f,\{u,d1\},\{v,d2\},\dots]} -- gives multiple
integral
\begin{equation}
\int_{\U(\text{\texttt{d1}})} dU  \int_{\U(\text{\texttt{d2}})} dV \dots
\text{\texttt{f}} .
\end{equation}
This function is presented in the example described in Section
\ref{ex:mean}.

 \item \texttt{IntegrateUnitaryHaarIndices[\{I1,J1,I2,J2\},dim]} -- 
 calculates the integral in Eq.~(\ref{intFormula}) for given multi-indices
 \texttt{I1,J1,I2,J2} and the dimension \texttt{dim} of the unitary group.
This function is presented in the example described in Section
\ref{ex:indices}.
\end{itemize}

\subsection{Weingarten function}

The main functions implemented in the  package, \texttt{IntegrateUnitaryHaar} and
\texttt{IntegrateUnitaryHaarIndices}, utilize the following functions to find
the value of the integral.
\begin{itemize}
 \item \texttt{Weingarten[type,dim]} -- returns the value of the Weingarten
  function given in Eq.~(\ref{eqn:weingarten-type-definition}) and accepts
  the following arguments
  \begin{itemize}
   \item \texttt{type} -- an integer partition which corresponds to cycle type 	
   of permutation (see Section \ref{sec:weingarten}),    
   \item \texttt{dim} -- the dimension of a unitary group, must be a positive 
   integer.
  \end{itemize}

\item \texttt{CharacterSymmetricGroup[part,type]} --  gives the character of 
the symmetric group $\chi^{\text{\texttt{part}}}(\text{\texttt{type}})$ (see
Section \ref{sec:characterSn}).

Parameter \texttt{type} is optional. The default value is set to a trivial
partition and in this case the function returns the dimension of the irreducible 
representation of symmetric group indexed by \texttt{part},  given by
Eq.~(\ref{eqn:hook}).   If \texttt{type} is specified the value of the character
is calculated by Murnaghan-Nakayama rule using  \texttt{MNInner} algorithm
provided in~\cite{bernstein2004computational}.

\item \texttt{SchurPolynomialAt1[part,dim]} -- returns the value of the Schur
polynomial $s_{\text{\texttt{part}}}$ at point  $(\underbrace{1,1, \dots
,1}_{\text{\texttt{dim}}})$, \ie{} the dimension of irreducible representation
of $\U(\text{\texttt{dim}})$ corresponding to  \texttt{part}, see
Eq.~(\ref{eqn:schurAt1}).
\end{itemize}

\subsection{Helper functions}
\begin{itemize}
\item \texttt{PermutationTypePartition[perm]} --
gives the partition which represents the cycle type of the permutation
\texttt{perm} (see Section \ref{sec:basic}).

\item \texttt{MultinomialBeta[p]} - 
for a given $d$-dimensional vector of non-negative numbers $p_1, p_2, \dots , p_d$
returns the value of multinomial Beta function defined as 
\begin{equation}
B(p) = 
\frac{\prod_{i=1}^d \Gamma(p_i) }
     {\Gamma(\sum_{i=1}^d p_i )}.
\end{equation}
This function is used in the optimization described in Section
\ref{sec:special-1-row}.

\item \texttt{ConjugatePartition[part]} --  gives a conjugate of a partition
\texttt{part} (see \cite{fultonharris}). This function is used for calculating
hook-length formula given by Eq.~(\ref{eqn:hook}).

\item \texttt{CardinalityConjugacyClassPartition[part]} -- gives  a cardinality
of a conjugacy class for the permutation with  cycle type given by partition
\texttt{part} (see \cite[Eq. 1.2]{sagan2001symmetric}). This function is used in
the optimization described in Section \ref{sec:even-powers}.

\item \texttt{BinaryPartition[part]} -- gives a binary representation of a
partition \texttt{part}. This function is needed for the implementation of 
\texttt{MNInner} algorithm.

\end{itemize}

\section{Usage examples} \label{sec:examples}
In order to present the main features of the described package we provide a 
series of examples.

\subsection{Elementary integrals} \label{ex:elementary}
Let us assume that $d=3$. We want to calculate the following integrals
\begin{eqnarray}
&& \int_{\U(d)} |u_{1,1}|^2 dU,\\
&& \int_{\U(d)}|u_{1,1}|^2|u_{2,2}|^2 dU, \\
&& \int_{\U(d)}u_{1,1} u_{2,2} \overline{u}_{1,2} \overline{u}_{2,1} dU .
\end{eqnarray}

Let us start by initializing the package
\IncludePdfExample{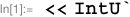}  
ow we calculate the integrals. 
\IncludePdfExample{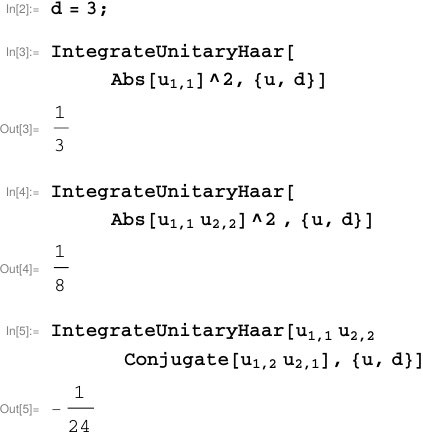}

\subsection{Operations on indices}\label{ex:indices}
Let us take the following set of multi-indices 
\begin{eqnarray}
I  = \{1, 1, 1, 2, 2\}, & & J  = \{2, 2, 1, 1, 1\} \\
I' = \{1, 1, 1, 2, 2\}, & & J' = \{2, 1, 1, 2, 1\}
\end{eqnarray}
and set $d = 6$. The above is equivalent to expression 
\begin{equation}
u_{1,2} u_{1,2} u_{1,1} u_{2,1} u_{2,1}
\overline{u}_{1,2} \overline{u}_{1,1} \overline{u}_{1,1} \overline{u}_{2,2} \overline{u}_{2,1}
\end{equation}
with symbolic variable $u$, which we aim to integrate over $\U(d)$.
After simplification the expression is equal to
\begin{equation}
|u_{1,1}|^2 |u_{1,2}|^2 |u_{2,1}|^2 u_{1,2} u_{2,1} \overline{u}_{1,1}
\overline{u}_{2,2}.
\end{equation}

After initializing the package and defining appropriate indices 
\IncludePdfExample{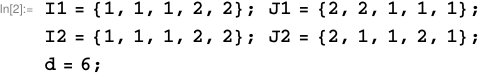}
we calculate the integral using provided function \texttt{IntegrateUnitaryHaarIndices} as 
\IncludePdfExample{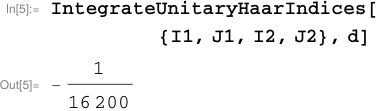}
which is equivalent to
\IncludePdfExample{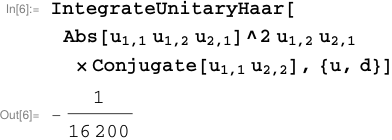}

\subsection{Matrix expressions}\label{ex:matrix}
\IntU{} package allows us to integrate matrix expressions, 
for example let us take $d=2$ and integrate 
\begin{eqnarray} \label{eqn:ex2}
 \int_{\U(d)} U^{\otimes 2} \otimes \bar{U} ^{\otimes 2} dU .
\end{eqnarray}

We define symbolic matrices $U \in \U(d)$ and $U2 = U \otimes U \in \U(d^2)$ as
\IncludePdfExample{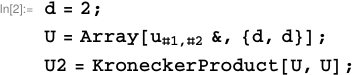}
and construct the integrand as
\IncludePdfExample{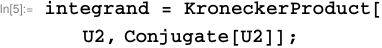} 
By using \texttt{IntegrateUnitaryHaar} function 
\IncludePdfExample{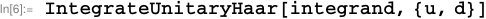}
we learn that the integral in Eq.~(\ref{eqn:ex2}) is equal to
\begin{equation}
\left(
\begin{smallmatrix}
 \frac{1}{3} & 0 & 0 & 0 & 0 & \frac{1}{6} & \frac{1}{6} & 0 & 0 &
   \frac{1}{6} & \frac{1}{6} & 0 & 0 & 0 & 0 & \frac{1}{3} \\
 0 & 0 & 0 & 0 & 0 & 0 & 0 & 0 & 0 & 0 & 0 & 0 & 0 & 0 & 0 & 0 \\
 0 & 0 & 0 & 0 & 0 & 0 & 0 & 0 & 0 & 0 & 0 & 0 & 0 & 0 & 0 & 0 \\
 0 & 0 & 0 & 0 & 0 & 0 & 0 & 0 & 0 & 0 & 0 & 0 & 0 & 0 & 0 & 0 \\
 0 & 0 & 0 & 0 & 0 & 0 & 0 & 0 & 0 & 0 & 0 & 0 & 0 & 0 & 0 & 0 \\
 \frac{1}{6} & 0 & 0 & 0 & 0 & \frac{1}{3} & -\frac{1}{6} & 0 & 0 &
   -\frac{1}{6} & \frac{1}{3} & 0 & 0 & 0 & 0 & \frac{1}{6} \\
 \frac{1}{6} & 0 & 0 & 0 & 0 & -\frac{1}{6} & \frac{1}{3} & 0 & 0 &
   \frac{1}{3} & -\frac{1}{6} & 0 & 0 & 0 & 0 & \frac{1}{6} \\
 0 & 0 & 0 & 0 & 0 & 0 & 0 & 0 & 0 & 0 & 0 & 0 & 0 & 0 & 0 & 0 \\
 0 & 0 & 0 & 0 & 0 & 0 & 0 & 0 & 0 & 0 & 0 & 0 & 0 & 0 & 0 & 0 \\
 \frac{1}{6} & 0 & 0 & 0 & 0 & -\frac{1}{6} & \frac{1}{3} & 0 & 0 &
   \frac{1}{3} & -\frac{1}{6} & 0 & 0 & 0 & 0 & \frac{1}{6} \\
 \frac{1}{6} & 0 & 0 & 0 & 0 & \frac{1}{3} & -\frac{1}{6} & 0 & 0 &
   -\frac{1}{6} & \frac{1}{3} & 0 & 0 & 0 & 0 & \frac{1}{6} \\
 0 & 0 & 0 & 0 & 0 & 0 & 0 & 0 & 0 & 0 & 0 & 0 & 0 & 0 & 0 & 0 \\
 0 & 0 & 0 & 0 & 0 & 0 & 0 & 0 & 0 & 0 & 0 & 0 & 0 & 0 & 0 & 0 \\
 0 & 0 & 0 & 0 & 0 & 0 & 0 & 0 & 0 & 0 & 0 & 0 & 0 & 0 & 0 & 0 \\
 0 & 0 & 0 & 0 & 0 & 0 & 0 & 0 & 0 & 0 & 0 & 0 & 0 & 0 & 0 & 0 \\
 \frac{1}{3} & 0 & 0 & 0 & 0 & \frac{1}{6} & \frac{1}{6} & 0 & 0 &
   \frac{1}{6} & \frac{1}{6} & 0 & 0 & 0 & 0 & \frac{1}{3}
\end{smallmatrix}
\right) .
\end{equation}

\subsection{Mean value of local unitary orbit}\label{ex:mean}
In this example we calculate the mean value of a local unitary orbit of a~given 
matrix $X \in {M}_{d^2}$
\begin{equation}\label{eqn:ex3}
 \Expectation [(U \otimes V) X (U \otimes V)^{\dagger}]. 
\end{equation}
We assume that $U$ and $V$ are stochastically independent random unitary
matrices of size 
$d$ distributed with the Haar measure. In this case we have
\begin{equation}
 \Expectation [(U \otimes V) X (U \otimes V)^{\dagger}]
 = \int_{\U(d)} \int_{\U(d)} (U \otimes V) X (U \otimes V)^{\dagger} dU dV.
\end{equation}
In this example we take $d = 3$.

We define symbolic matrices $X$ of size $d^2$ and $U,V \in \U(d)$ as 
\IncludePdfExample{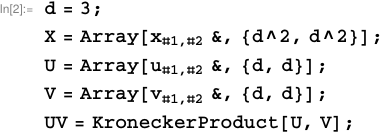} 
Using \texttt{IntegrateUnitaryHaar} function with two variable specifications 
we calculate the double integral
\IncludePdfExample{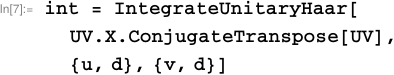}
and find out that the expectation value in Eq.~(\ref{eqn:ex3}) is equal to
\begin{equation}
 \Expectation [(U \otimes V) X (U \otimes V)^{\dagger} ]
 =  
 \frac{1}{d^2} \tr X \ \1_{d\times d}.
\end{equation}

Similarly one can calculate the covariance tensor of the local unitary orbit of 
a~given matrix $X \in {M}_{d^2}$. If
$
z = (U \otimes V) X (U \otimes V)^{\dagger}
$
then the covariance tensor is given by \cite{DGHMPZ12}
\begin{equation}
 \Expectation[\{z_{ij} \overline{z}_{kl}\}_{ijkl}] = 
 \Expectation[(U \otimes V) X (U \otimes V)^{\dagger} \otimes 
 \overline{(U\otimes V) X (U \otimes V)^{\dagger}}] .
\end{equation}
Using \texttt{IntegrateUnitaryHaar} one can check that for the fixed
dimension the integral agrees with the calculations presented in~\cite{DGHMPZ12}.

\subsection{Moments of maximally entangled numerical shadow} \label{ex:moments}
If $U$ is a random $d \times d$ unitary matrix distributed according to the Haar
measure  \cite{miszczak12generating} then the pure state obtained by vectorization  $| \xi \rangle  =
\frac{1}{ \sqrt{d} } \mathrm{vec}(U)$ is maximally entangled  on ${H}_A
\otimes {H}_B = \mathbb{C}^d \otimes \mathbb{C}^d$. Moreover, state $|
\xi \rangle $  has a distribution invariant to multiplication by local unitary
matrices.  We denote the corresponding probability measure on pure states of
size $d \times d$ as $\mu$. The numerical shadow of operator $X \in
{M}_{d^2}$ with respect to this measure (maximally entangled numerical
shadow) is defined  as
\begin{equation}
P_X(z) = \int_{\mathbb{C}^{d^2}}  \delta(z - \langle \psi| X |\psi\rangle) \
\mathrm{d}\mu(\psi).
\end{equation}
The corresponding probability measure is denoted by $ d\mu^{\epsilon }_{X}$. For
definition and basic facts concerning numerical shadows see
\cite{DGHMPZ12,dunkl11shadow1,dunkl11shadow2}.

The first two moments of $ d\mu^{\epsilon}_{X}$ are calculated in 
\cite{DGHMPZ12},
and are given by 
\begin{equation} \label{eqn:max-ent-mean}
\int_{\mathbb{C}}z \; d\mu^{\epsilon}_{X}\left(  z\right)  
=\frac{1}{d^{2}}\mathrm{tr}X \; ,
\end{equation}
and 
\begin{eqnarray} \label{eqn:max-ent-sec-mom}
\int_{\mathbb{C}}z\overline{z}\; d\mu^{\epsilon}_{X} \left(z\right)  
 &=&
 \frac{1}{d^2(d^2-1)} 
 \left( 
   |\tr X|^2 
   +  
  \|X\|_{\mathrm{HS}}^2 
 \right)
 \\\nonumber
 &-&
 \frac{1}{d^3(d^2-1)}  \left(
  \|\tr_A (X)\|_{\mathrm{HS}}^2  
  + 
  \|\tr_B (X)\|_{\mathrm{HS}}^2 
 \right),
\end{eqnarray}
where $\tr_{A}$ and $\tr_{B}$ denote partial traces over a specified sub-system
and $\|\cdot\|_{\mathrm{HS}}$ is a Hilbert-Schmidt norm given by
$\|X\|_{\mathrm{HS}} = \sqrt{\tr X X^{\dagger}}$. 

In order to calculate (\ref{eqn:max-ent-sec-mom}) and (\ref{eqn:max-ent-sec-mom}) 
define symbolic matrices
\IncludePdfExample{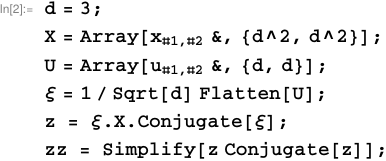} 
Now we calculate the first moment
\IncludePdfExample{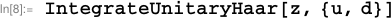}
and the second moment 
\IncludePdfExample{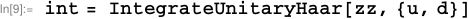} 
After some algebraic manipulations one can
see that the above agrees with Eqs.~(\ref{eqn:max-ent-mean}) and (\ref{eqn:max-ent-sec-mom}).

Package \IntU{} has been used successfully  in the context of quantum 
entanglement in paper~\cite{enriquez2015minimal}, where authors use it to 
calculate the moments of the three-tangle.

\section{Summary}\label{sec:summary}
We described \IntU{} package for \Mathematica{} computing system for calculating
 polynomial integrals over $\U(d)$ with respect to Haar measure. We described 
some number of special cases which can be used to optimize the calculation speed
 for some classes of integrals. We also provide some examples of the usage of
the presented package including the applications in the geometry of the quantum
states.

Calculation time of the package strongly depends on a degree of the integrand.
For polynomials of small degree the package is able to calculate the value of
integral using the direct formula (\ref{intFormula}). For polynomials of large
degree the calculation time grows rapidly and the calculation  is possible only
if one of the special cases (optimizations) is used.

Nevertheless, the presented package can be very useful in the investigations
involving Circular Unitary Ensemble and the geometry of quantum states and 
quantum entanglement.

\section{Acknowledgements}   
Work of Z.~Pucha{\l}a was partially supported by the Polish National Science
Centre under the research project N~N514~513340 while 
work of J.~A.~Miszczak was partially supported by the Polish National Science 
Centre under the research project N N516 475440. Authors would like to thank 
K.~\.Zyczkowski and P.~Gawron for motivation and interesting discussions.

\end{document}